\documentclass[a4paper,floatfix,prd,superscriptaddress,reprint,11pt]{article}
\usepackage[utf8]{inputenc}
\usepackage[pdftex]{graphicx}
\usepackage{pos}
\usepackage{fontenc}
\usepackage{times}
\usepackage{graphicx}
\usepackage{appendix}
\usepackage{amsmath}
\usepackage{dsfont}
\usepackage{array}
\usepackage{bm}
\usepackage{multirow}
\usepackage{upgreek}
\usepackage{xcolor}
\usepackage{verbatim}
\usepackage{slashed}
\usepackage{cleveref}
\usepackage{caption}
\usepackage{subcaption}
\usepackage{mathtools}
\usepackage{url}

\usepackage[braket,qm]{qcircuit}

\title{Mass Renormalization of the Schwinger Model with Wilson and Staggered Fermions in the Hamiltonian Lattice Formulation}
\ShortTitle{Mass Renormalization of the Schwinger Model}

\author*[a, b]{Takis Angelides}
\author[c,d]{Lena Funcke}
\author[b]{Karl Jansen}
\author[b,e]{Stefan Kühn}

\affiliation[a]{Institut für Physik, Humboldt-Universität zu Berlin, Newtonstr. 15, 12489 Berlin, Germany}

\affiliation[b]{Deutsches Elektronen-Synchrotron DESY, Platanenallee 6, 15738 Zeuthen, Germany}

\affiliation[c]{Transdisciplinary Research Area ``Building Blocks of Matter and Fundamental Interactions'' (TRA Matter) and Helmholtz Institute for Radiation and Nuclear Physics (HISKP), University of Bonn, Nußallee 14-16, 53115 Bonn, Germany}

\affiliation[d]{Center for Theoretical Physics, Co-Design Center for Quantum Advantage, and NSF AI Institute for Artificial Intelligence and Fundamental Interactions, Massachusetts Institute of Technology, 77 Massachusetts Avenue, Cambridge, MA 02139, USA}

\affiliation[e]{Computation-Based Science and Technology Research Center, The Cyprus Institute, 20 Kavafi Street,
2121 Nicosia, Cyprus}

\emailAdd{takis.angelides@desy.de}
\emailAdd{lfuncke@uni-bonn.de}
\emailAdd{karl.jansen@desy.de}
\emailAdd{stefan.kuehn@desy.de}

\abstract{Lattice computations in the Hamiltonian formulation have so far mainly focused on staggered fermions. In these proceedings, we study Wilson fermions in the Hamiltonian formulation and propose a new method to determine the resulting mass shift. As a benchmark study, we examine the one-flavour Schwinger model with Wilson fermions and a topological $\theta$-term using matrix product states. Wilson fermions explicitly break chiral symmetry; thus, the bare mass of the lattice model receives an additive renormalization. In order to measure this mass shift directly, we develop a method that is suitable for the Hamiltonian formulation, which relies on the fact that the vacuum expectation value of the electric field density vanishes when the renormalized mass is zero. We examine the dependence of the mass shift on the lattice spacing, the lattice volume, the $\theta$-parameter, and the Wilson parameter. Using the mass shift, we then perform the continuum extrapolation of the electric field density and compare the resulting mass dependence to the analytical predictions of mass perturbation theory. We demonstrate that incorporating the mass shift significantly improves the continuum extrapolation. Finally, we apply our method to the same model using staggered fermions instead of Wilson fermions and compare the resulting mass shift to recent theoretical predictions.} 

\FullConference{The 39th International Symposium on Lattice Field Theory, (Lattice 2022)\\
  8-13 August 2022\\
  Hörsaalzentrum Poppelsdorf, Bonn, Germany
}

\begin{document}
\maketitle

\section{Introduction}

The Kogut-Susskind staggered formulation of lattice fermions~\cite{kogut_susskind_staggered} has so far been the most common choice in studies of lattice gauge theories in the Hamiltonian formulation. An alternative approach are Wilson fermions~\cite{wilson_fermions_original_work}, for which the doublers acquire a mass proportional to the inverse lattice spacing and thus are completely removed when the lattice spacing is taken to zero. The implementation of Wilson fermions generalises trivially to any number of spatial dimensions. However, Wilson fermions break chiral symmetry explicitly, which generates an additive mass renormalization. The knowledge of this mass shift allows for a direct examination of the mass dependence of observables and for following a massless renormalization scheme. However, this shift has so far been only determined for lattice gauge theories in the Lagrangian formulation.
\par In these proceedings, we propose a method to determine the mass shift of Wilson fermions in the Hamiltonian formulation. As a benchmark model, we use the Hamiltonian formulation of the Schwinger model in the presence of a topological $\theta$-term~\cite{coleman_more_about_SM}. This model describes quantum electrodynamics in 1+1 dimensions, and the $\theta$-term is related to a background electric field~\cite{tong_gt}. The $\theta$-term induces a sign problem~\cite{wiese_troyer_sign_problem} for standard lattice Monte Carlo simulations; we therefore use the tensor-network technique of matrix product states (MPS)~\cite{practical_intro_mps}, which circumvents this problem.
\par The rest of this article is organized as follows. In Sec.~\ref{theory_section}, we provide the theoretical background of the Schwinger model both in the continuum and on the lattice. Section \ref{methods_section} describes the methods used, including MPS and the new method to determine the mass shift. Our results are presented in Sec.~\ref{results_section} before concluding in Sec.~\ref{conclusion_section}. Finally, we note that our method of computing the mass shift is not restricted to a particular type of lattice fermions. Hence, while the main text focuses on Wilson fermions, in Appendix \ref{staggered_appendix} we consider staggered fermions and compute their mass shift in the Schwinger model to compare it to the recent theoretical predictions in Ref.~\cite{chiral_dempsey_staggered}.

\section{Theory}
\label{theory_section}

In this section, we first review the continuum Schwinger model, including the analytical prediction for the vacuum expectation value of the electric field density. We then describe the lattice implementation of Wilson fermions and provide the final lattice Hamiltonian.

\subsection{Continuum Theory}

In the continuum, the massive Schwinger model~\cite{lagrangian_of_schwinger_model} for a single fermion flavour including a topological $\theta$-term can be described by the Hamiltonian density
\begin{equation}
\label{hamiltonian_density}
    \mathcal{H} = -i\overline{\psi}\gamma^1\left(\partial_1 - igA_1\right)\psi + m\overline{\psi}\psi + \frac{1}{2}\left(\dot{A}_1 + \frac{g\theta}{2\pi}\right)^2,
\end{equation}
where the fermionic field $\psi$ is a two-component spinor in 1+1 dimensions and $m$ corresponds to its bare mass. This Hamiltonian is in the temporal gauge, $A_0 = 0$, hence the gauge field appears only through $F_{01} = \dot{A}_1$. Further, $g$ is the coupling constant between the fermions and the gauge field, which has units of mass and does not renormalize since the model is super-renormalizable~\cite{super_renormalizability, adam}. The $\theta$-term corresponds to a constant background electric field given by $g\theta / 2\pi$~\cite{tong_gt, adam, coleman_more_about_SM,topo_vac}. The massless Schwinger model has an axial anomaly, which implies that the classical axial symmetry $\psi \to e^{i\gamma_5\alpha}\psi$ is no longer a symmetry on the quantum level. Consequently, the $\theta$-parameter is unphysical in the massless case; hence, observables should not depend on $\theta$~\cite{manton, topo_vac}.
\par The observable we consider in our numerical study is the vacuum expectation value of the electric field density, which was analytically derived for the continuum model within mass perturbation theory up to order $\mathcal{O}((m/g)^2)$ in Ref.~\cite{adam},
\begin{equation}
\label{adam_efd}
    \frac{\mathcal{F}}{g} = \frac{e^\gamma}{\sqrt{\pi}}\left(\frac{m}{g}\right)\sin{(\theta)} - 8.9139\text{ }\frac{e^{2\gamma}}{4\pi}\left(\frac{m}{g}\right)^2\sin{(2\theta)},
\end{equation}
where $\gamma = 0.5772156649$ is the Euler-Mascheroni constant. 

\subsection{Lattice Theory}
\label{lattice_theory}

Following Refs.~\cite{zache, tavernelli_wilson_fermions}, we now derive the lattice version of the Schwinger model with Wilson fermions, which yields Eq.~\eqref{hamiltonian_density} in the continuum. We start from the free Dirac Hamiltonian,
\begin{equation}
    H^{\text{free}} = \int dx \text{ } \overline{\psi}\left(-i\gamma^1\partial_1 + m\right)\psi,
\end{equation}
and use the symmetric lattice derivative, $\partial_1 \psi_n = (\psi_{n+1}-\psi_{n-1})/2a$, along with $\int dx \to a\sum_n$, where $a$ is the lattice spacing and $n$ is the lattice site index, to arrive at the discretized Hamiltonian
\begin{equation}
\label{discrete_free_hamiltonian}
    H^\text{free}_\text{lat} = a\sum_n m_\text{lat}\overline{\psi}_n\psi_n -i\overline{\psi}_n \gamma^1\left(\frac{\psi_{n+1}-\psi_{n-1}}{2a}\right).
\end{equation}
In Eq.~\eqref{discrete_free_hamiltonian}, we distinguish the lattice mass $m_\text{lat}$ from the continuum mass $m$. To implement Wilson fermions~\cite{wilson_fermions_original_work}, we add to Eq.~\eqref{discrete_free_hamiltonian} the Wilson term
\begin{equation}
\label{wilson_term}
    \Delta H_{\rm Wilson} = -r\frac{a^2}{2}\sum_n\overline{\psi}_n\partial_1^2\psi_n = -r\frac{a^2}{2}\sum_n\overline{\psi}_n\left(\frac{\psi_{n+1}+\psi_{n-1}-2\psi_n}{a^2}\right),
\end{equation}
where $r$ is the Wilson parameter, and we used the discrete symmetric second derivative. The resulting mass of the doublers is proportional to $r$~\cite{wilson_fermions_original_work}, and for our simulations we make the standard choice of $r = 1$. The free lattice Dirac Hamiltonian with Wilson fermions thus reads
\begin{equation}
\label{free_lattice_wilson_hamiltonian}
    H^\text{free}_\text{lat, Wilson} =\sum_n(am_\text{lat} + r)\overline{\psi}_n\psi_n-\overline{\psi}_n\left(\frac{r+i\gamma^1}{2}\right)\psi_{n+1}+\overline{\psi}_n\left(\frac{-r+i\gamma^1}{2}\right)\psi_{n-1}.
\end{equation}
Equation~\eqref{free_lattice_wilson_hamiltonian} has a global $U(1)$ symmetry, $\psi_n \to e^{i\alpha}\psi_n$, and gauging the theory allows for invariance under local gauge transformations, $\psi_n \to e^{i\alpha_n}\psi_n$ and $U_n \to e^{i\alpha_n}U_ne^{-i\alpha_{n+1}}$. Here, the link operator $U_n$ is placed on the link between sites $n$ and $n+1$. The electric field operator $E_n$ is conjugate to $U_n$, which implies $[E_{n} , U_{n'}] = g \delta_{n,n'} U_{n'}$. Thus, the resulting interacting lattice Hamiltonian is
\begin{equation}
\label{interacting_discrete_hamiltonian}
     H^\text{int}_\text{lat, Wilson} = \sum_n-\overline{\psi}_n\left(\frac{r+i\gamma^1}{2}\right)U_n\psi_{n+1} + \overline{\psi}_n\left(\frac{-r+i\gamma^1}{2}\right)U^\dagger_n\psi_{n-1} + (am_\text{lat}+r)\overline{\psi}_n\psi_n + a\frac{E_n^2}{2}.
\end{equation}
In the Hamiltonian lattice formulation, it is convenient to use dimensionless quantities in units of the coupling $g$. The dimensionless form of the field operators is given by $L_n = E_n/g$ and $\phi_{n, \alpha} = (-1)^n\sqrt{a}\psi_{n, \alpha}$. The physical volume in units of the coupling is given by $agN = N/\sqrt{x}$, where we defined the dimensionless inverse lattice spacing squared $x \equiv 1/(ag)^2$. 
\par The physical part of the Hilbert space is defined as the set of states which obey Gauss's law, $L_n - L_{n-1} = Q_n$, with the charge operator defined as $Q_n \equiv \psi^\dagger_n\psi_n - 1$. We use open boundary conditions (OBC) and hence can solve this constraint explicitly after fixing the left boundary link to a given input value, $l_0 \equiv \theta / 2\pi$. This parameter $l_0$ is precisely the dimensionless background electric field. After fixing $l_0$, we can express the dimensionless electric field operator as
\begin{equation}
    L_n = l_0 + \sum_{k=1}^{n}Q_k.
    \label{l_n_equation}
\end{equation}
To enforce Gauss's law, we can then substitute the expression for $L_n$ in Eq.~\eqref{l_n_equation} into our Hamiltonian in Eq.~\eqref{interacting_discrete_hamiltonian}. Furthermore, the gauge degrees of freedom  in Eq.~\eqref{interacting_discrete_hamiltonian} can be completely eliminated by applying a unitary transformation, $\phi_n \to \prod_{k < n}U^\dagger_k \phi_n$~\cite{hamer97}.
Using the mapping $\phi_{n,\alpha} \to \chi_{2n-2+\alpha}$ allows us then to translate our operators to Pauli spin operators using the Jordan-Wigner transformation~\cite{jordan_wigner}, $\chi_n = \prod_{k<n}(i\sigma^z_k)\sigma_n^-$, where the ladder Pauli matrices are $\sigma_n^{\pm} = \frac{1}{2}(\sigma_n^x\pm i\sigma_n^y)$.
\par Finally, we make the Hamiltonian dimensionless, $\widetilde{W} = (2/ ag^2)H^\text{int}_\text{lat, Wilson}$, and add a penalty term, $W = \widetilde{W} + \lambda \left(\sum_{n=1}^NQ_n\right)^2$, with $\lambda$ set to 100. This term enforces zero total charge by increasing the energy of the states with nonzero total charge. The final dimensionless Hamiltonian reads
\begin{align}
\label{model_hamiltonian_w}
\begin{split}
    W = &\text{ }ix(r-1)\sum_{n=1}^{N-1}\left(\sigma_{2n}^-\sigma_{2n+1}^+ - \sigma_{2n}^+\sigma_{2n+1}^-\right) + \lambda \left(\sum_{n=1}^NQ_n\right)^2 \\
    &+ix(r+1)\sum_{n=1}^{N-1}\left(\sigma_{2n-1}^+\sigma^z_{2n}\sigma^z_{2n+1}\sigma^-_{2n+2} - \sigma_{2n-1}^-\sigma^z_{2n}\sigma^z_{2n+1}\sigma^+_{2n+2}\right) \\
    &+\sum_{n=1}^{N-1}\left(l_0 + \sum_{k=1}^n Q_k\right)^2 + 2i\left(\frac{m_\text{lat}}{g}\sqrt{x} + xr\right)\sum_{n=1}^N\left(\sigma_{2n-1}^-\sigma_{2n}^+ - \sigma_{2n-1}^+\sigma_{2n}^-\right).
    \end{split}
\end{align}

\section{Methods}
\label{methods_section}
In this section, we first review the MPS algorithm used to find the ground state of the Hamiltonian in Eq.~\eqref{model_hamiltonian_w}. Then, we introduce the new method to determine the mass shift by measuring the mass dependence of the vacuum expectation value of the electric field density.

\subsection{Matrix Product States}
\label{sec:MPS}
Matrix product states are a parametric ansatz composed of tensors, with the tensor size limiting the maximum entanglement that can be present in the state~\cite{practical_intro_mps}. For OBC, the ansatz is given by
\begin{equation}
\label{mps_ket_rep}
    \ket{\psi} = \sum_{i_1,i_2,..,i_{2N}} A^{i_1}_{1,\alpha_1}A^{i_2}_{\alpha_1,\alpha_2}...A^{i_{2N}}_{\alpha_{2N-1},1} \ket{i_1} \otimes \ket{i_2} ... \otimes \ket{i_{2N}},
\end{equation}
with implicit contraction of the indices $\alpha_k$. The maximum value of these contracted indices is called the bond dimension $D$. The physical indices $i_n$ range over the physical degrees of freedom on each site $n$ and---in our case---can take two possible values corresponding to spin up or down, since the states $\ket{i_n}$ correspond to spin states. The $A$-tensors are the building blocks of the MPS, which are rank-3 tensors that, upon contraction via the bond indices, form an approximation to the rank-$(2N)$ tensor of the quantum state. To compute the ground state of the Hamiltonian $W$, we employ a variational algorithm~\cite{schollwock} to minimize the energy $E=\langle\psi|W|\psi\rangle$ using the ITensors Julia library~\cite{iTJulia}. The algorithm optimizes the tensors until the relative change in the energy is below a certain accuracy $\eta$, which we set to $10^{-12}$. Subsequently, we can use the MPS result for the ground state $|\psi_0\rangle$ to perform measurements of observables, $\langle\psi_0|O|\psi_0\rangle$. 

\subsection{Method to Compute the Mass Shift}
\label{measurements_section}

\par Our goal is to determine the mass shift that arises due to the Wilson term in Eq.~\eqref{wilson_term}. For this, we express the renormalized mass $m_r/g$ in terms of the lattice mass $m_{\text{lat}}/g$,
\begin{equation}
\label{mass_shift_general_equation}
    \frac{m_r}{g} = \frac{m_{\text{lat}}}{g} + \text{MS},
\end{equation}
where the term MS refers to the mass shift in units of the coupling $g$.

To compute this mass shift, we exploit the fact that the vacuum expectation value of the dimensionless electric field density $\mathcal{F}/g$ vanishes when the renormalized mass $m_r$ is zero. Thus, we measure $\mathcal{F}/g$ as a function of the lattice mass $m_\text{lat}/g$, as shown in Fig.~\ref{efd_vs_mlat_and_ms_vs_inv_vol}(a). Guided by the analytical continuum prediction in Eq.~\eqref{adam_efd}, we then fit a quadratic function to our data and identify the term MS in Eq.~\eqref{mass_shift_general_equation} with minus the value of $m_\text{lat}/g$ for which $\mathcal{F}/g = 0$. We note that this method requires simulations at negative values of $m_\text{lat}/g$ and thus is not applicable to standard lattice Monte Carlo simulations in the Lagrangian formulation due to the sign problem.

\par To measure $\mathcal{F}/g$ using the MPS algorithm explained in Sec.~\ref{sec:MPS}, we first use Eq.~\eqref{l_n_equation} to calculate $L_n = E_n/g$. We only keep links from the center of the lattice to reduce boundary effects that might be present due to using OBC. The value measured for $\mathcal{F}/g$ is then the mean value of the electric field on those links. Errors arising from the extrapolation of $\mathcal{F}/g$ to infinite bond dimension $D$ and to infinite inverse lattice spacing $x$ are added in quadrature according to Appendix B of Ref.~\cite{topo_vac}. Unlike in Ref.~\cite{topo_vac}, we explicitly work at fixed finite volume $N/\sqrt{x}$ and therefore extrapolate only to infinite bond dimension $D$ and infinite $x$.

\section{Results}
\label{results_section}

\par In this section, we show our numerical results for the mass shift and its dependence on the lattice volume $N/\sqrt{x}$, the lattice spacing $ag$, the $\theta$-parameter, and the Wilson parameter $r$.
We present the qualitative behaviour of these dependencies and provide quantitative results where appropriate.

\par
\begin{figure}[t!]
     \centering
     \includegraphics[width=0.94\textwidth]{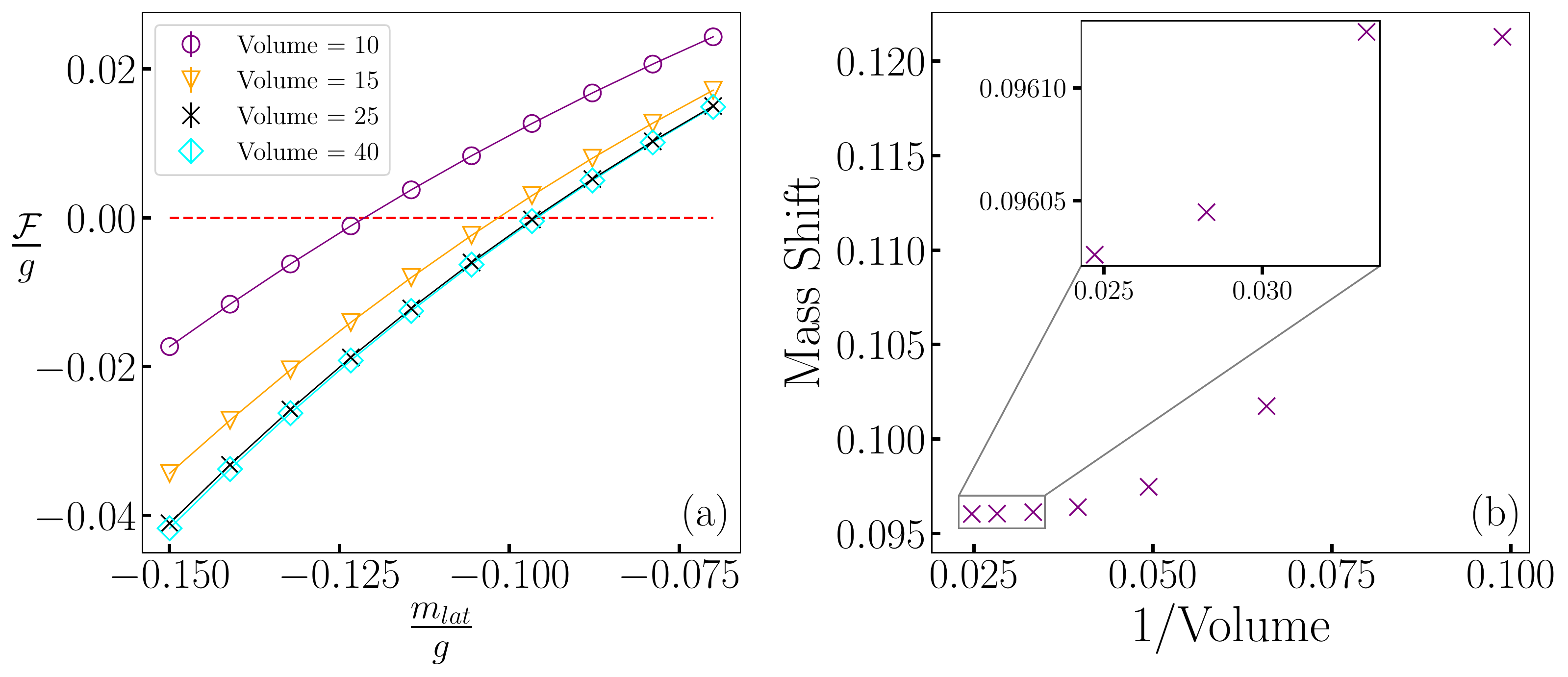}
     \caption{\textbf{(a)} Electric field density  $\mathcal{F}/g$ as a function of the lattice mass $m_\text{lat}/g$, for different volumes $N/\sqrt{x}$. Equation~\eqref{adam_efd} indicates that $\mathcal{F}/g$ vanishes at $m_r/g = 0$, therefore the intercept with $\mathcal{F}/g = 0$ (red dashed line) gives minus the mass shift. \textbf{(b)} Mass shift against inverse volume. The inset plot shows the three points with the largest volumes; the relative difference in the mass shift for the largest volumes 35 and 40 is $\sim 0.01$\%. For the data of both panels, we fix $x = 10$ and $l_0 = 0.1$ and consider volumes $N/\sqrt{x}$ between 10 and 40.}
     \label{efd_vs_mlat_and_ms_vs_inv_vol}
\end{figure}

 First, we observe that the mass shift strongly depends on the lattice volume $N/\sqrt{x}$, as shown in Fig.~\ref{efd_vs_mlat_and_ms_vs_inv_vol}(b). This is due to the presence of finite-size effects for the observable used to calculate the mass shift, namely the vacuum expectation value of the electric field density $\mathcal{F}/g$, as shown in Fig.~\ref{efd_vs_mlat_and_ms_vs_inv_vol}(a). However, this dependence becomes negligible for sufficiently large volumes, ${N}/{\sqrt{x}} \gtrsim 40$.

\begin{figure}[t!]
     \centering
     \includegraphics[width=0.94\textwidth]{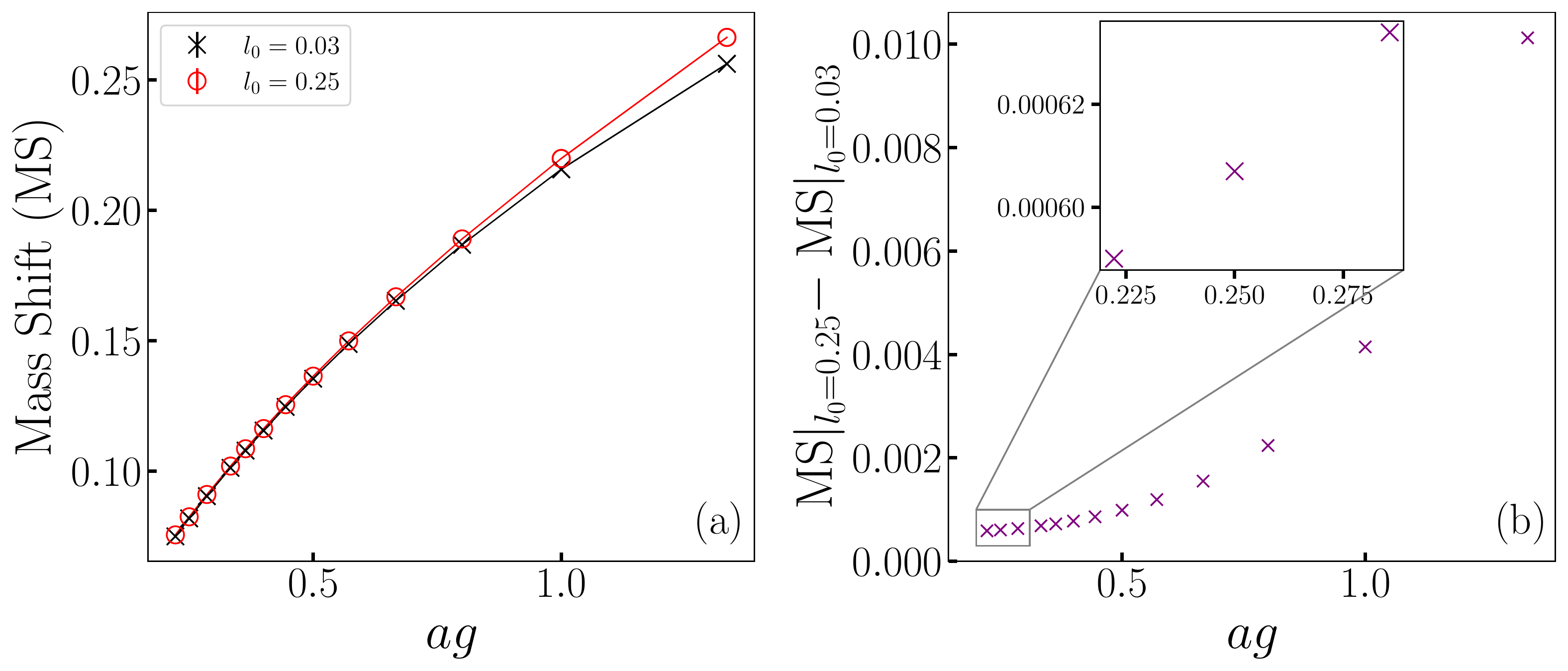}
     \caption{ \textbf{(a)} Mass shift (MS) versus lattice spacing $ag= 1/\sqrt{x}$ for two values of the background field, $l_0 = \theta/(2\pi) = 0.03$ and $0.25$. The volume $N/\sqrt{x} = 20$ is fixed, with $N$ ranging from 25 to 90. \textbf{(b)} Mass shift at $l_0 = 0.25$ minus mass shift at $l_0 = 0.03$ as a function of the lattice spacing $ag$.}
     \label{mass_shift_and_difference_in_mass_shift_vs_ag}
\end{figure}

Next, we examine the dependence of the mass shift on the lattice spacing $ag$ and on the \mbox{$\theta$-parameter}. As shown in Fig.~\ref{mass_shift_and_difference_in_mass_shift_vs_ag}(a), the mass shift strongly depends on the lattice spacing, as expected. Moreover, the mass shift slightly depends on $l_0=\theta/2\pi$, but this dependence is shown to vanish in Fig.~\ref{mass_shift_and_difference_in_mass_shift_vs_ag}(b) as $ag$ is taken to zero. Note that the mass shift only becomes strictly independent of $l_0$ once the infinite-volume limit is taken. This is also expected, because the continuum theory predicts that for $m=0$ the $\theta$-parameter is unphysical and should not affect any measurement, due to the axial anomaly. However, the axial anomaly on the lattice is not exact~\cite{topo_vac}, which causes the slight $\theta$-dependence of the mass shift for large lattice spacings.
\par Regarding the dependence of the mass shift on the Wilson parameter $r$, we numerically confirmed that the mass shift is antisymmetric in $r \to -r$. This is due to the spurious symmetry $\psi \to \gamma_5 \psi$, $m_\text{lat}/g \to -m_\text{lat}/g$, and $r \to -r$ of the Hamiltonian in Eq.~\eqref{interacting_discrete_hamiltonian}.

\par
\begin{figure}[t!]
     \centering
     \includegraphics[width=0.94\textwidth]{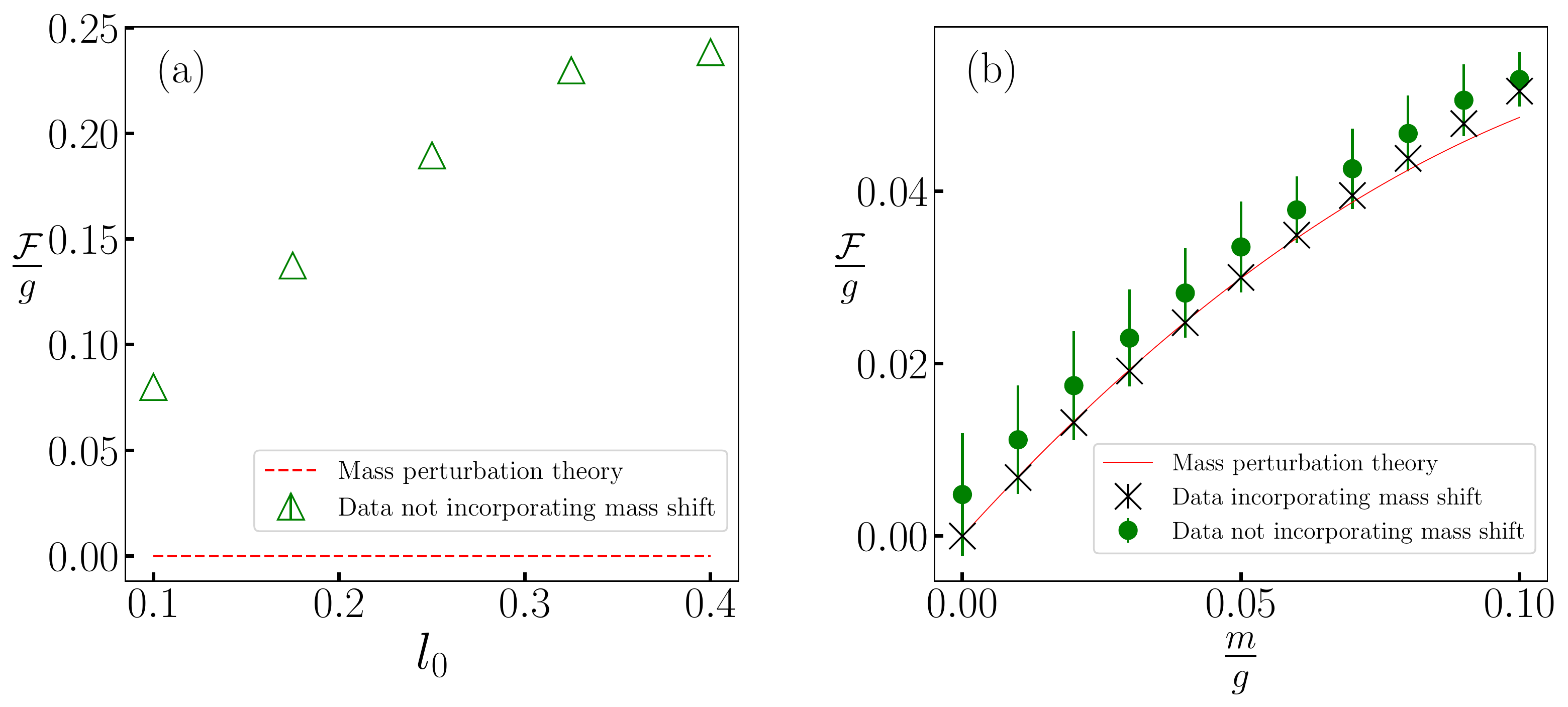}
     \caption{\textbf{(a)} Electric field density $\mathcal{F}/g$ as a function of $l_0 = \frac{\theta}{2\pi}$ for $m_{\rm lat}/g = 0$, $N = 32$, $x = 0.6$, $N/\sqrt{x} = 40$.\linebreak Equation~\eqref{adam_efd} indicates that $\mathcal{F}/g$ should be zero  at $m/g = 0$ for any $l_0$ (red dashed line). Without incorporating the mass shift, the data (green triangles) increasingly deviate from this expectation for increasing $l_0$, as the mass shift grows with $l_0$. \textbf{(b)} Continuum-extrapolated electric field density $\mathcal{F}/g$ as a function of $m/g$. We fixed the volume to $N/\sqrt{x} = 20$, the background field to $l_0 = \frac{\theta}{2\pi} = 0.125$, and extrapolated to the continuum limit $ag\to 0$ using eight points of $N$ between 100 and 300. The theoretical continuum prediction (red line) comes from mass perturbation theory, see Eq.~\eqref{adam_efd}. For small $m/g$, where the prediction is valid, the data incorporating the mass shift, $m/g = m_\text{lat}/g$ + MS, (black crosses) follow this result with excellent agreement and small error bars. The data without incorporating the mass shift, $m/g \equiv m_\text{lat}/g$,  (green circles) have larger error bars and mean values that are further away from the continuum result.}
     \label{advantage_of_mass_shift}
\end{figure}

\begin{figure}
     \centering
     \includegraphics[width=0.94\textwidth]{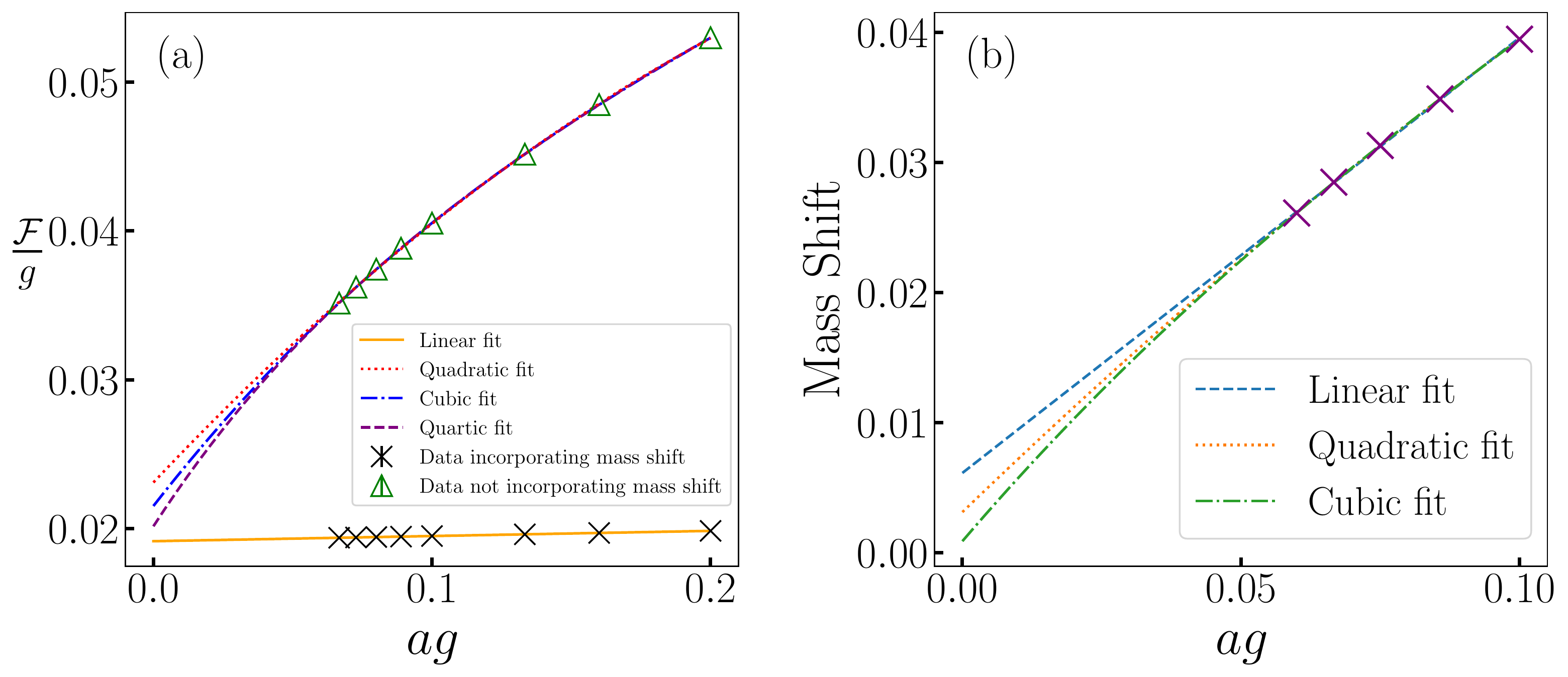}
     \caption{\textbf{(a)} Electric field density $\mathcal{F}/g$ versus lattice spacing $ag$ for $l_0 = 0.125$, $N/\sqrt{x} = 20$, and $m_r/g=0.03$ (black crosses) and $m_\text{lat}/g=0.03$ (green triangles). A linear fit is sufficient for the black cross data, as it is expected from non-improved Wilson fermions which have an $\mathcal{O}(ag)$ scaling behaviour~\cite{order_a_wislon}, whereas for the green triangles we used the following procedure. To approximate $\mathcal{F}/g$ at $ag\to 0$, a quadratic, cubic, and quartic polynomial is fitted and a weighted average of the resulting $y$-intercepts is taken with the weights corresponding to the mean square error of each fit. The error on the $y$-intercept for each fit is set according to Sec.~\ref{measurements_section}. The final error is then found by adding the individual errors in weighted quadrature. \textbf{(b)} Mass shift versus $ag$ for $N/\sqrt{x} = 30$, $l_0 = 0.125$ and $N$ ranging from 300 to 500. We have approximated the value at $ag = 0$ using the same fitting procedure as described for the green triangle points in \textbf{(a)}.}
     \label{efd_and_mass_shift_vs_ag}
\end{figure}

 Finally, we wish to demonstrate that incorporating the mass shift significantly improves the continuum extrapolation of the electric field density $\mathcal{F}/g$. Figure \ref{advantage_of_mass_shift}(a) shows that without incorporating the mass shift, the lattice data for $\mathcal{F}/g$ deviate from the expectation that $\mathcal{F}/g$ should vanish at $m/g = 0$ for any $l_0$. This deviation becomes more severe for larger $l_0$, because the mass shift grows with $l_0$. Using our method, we can utilize the condition $\mathcal{F}/g\equiv 0$ to compute the mass shift in lattice units (MS) [see Eq.~\eqref{mass_shift_general_equation}] for certain lattice parameters, thus allowing us to keep the renormalized mass $m_r/g$ constant while changing the lattice spacing. For the continuum extrapolation of $\mathcal{F}/g$ at fixed volume, as shown in Fig.~\ref{advantage_of_mass_shift}(b), the incorporation of the mass shift significantly improves the convergence towards the continuum prediction of mass perturbation theory, both in terms of decreasing the error bars and in terms of yielding mean values that are consistent with the prediction for small $m/g$. The large error bars for the data without the mass shift can be explained by Fig.\ \ref{efd_and_mass_shift_vs_ag}(a), which demonstrates that data incorporating the mass shift converge faster when taking the limit $ag\to 0$. Indeed, the data with the mass shift in Fig.\ \ref{efd_and_mass_shift_vs_ag}(a) already converge to the continuum value of $\mathcal{F}/g$ for large values of $ag$. In contrast, for the data without the mass shift, the continuum extrapolation for $\mathcal{F}/g$ strongly deviates from the lattice result obtained at the smallest lattice spacing. The mass shift is shown in Fig.\ \ref{efd_and_mass_shift_vs_ag}(b), which is linear in $ag$ to first order and after extrapolating to $ag \to 0$ yields $\text{MS} =0.016 \pm 0.009$ for a fixed volume of $N/\sqrt{x}$ = 30, which is consistent with a vanishing mass shift within two standard deviations. This value for MS would be even closer to zero if the infinite-volume limit is also taken, as can be inferred from Fig.\ \ref{efd_vs_mlat_and_ms_vs_inv_vol}(b).  

\par All in all, we can express the renormalized mass $m_r/g$ in terms of the lattice mass $m_\text{lat}/g$ using the parametric expression $m_r/g = m_\text{lat}/g + \text{MS}(N, x, \theta, r)$. In this section, we quantified the mass shift in lattice units, MS, and explored its behaviour as a function of each argument $N$, $x$, $\theta$, and $r$.

\section{Conclusion}
\label{conclusion_section}

In these proceedings, we developed a new method to determine the mass shift of Wilson fermions in the Hamiltonian formulation. Using the Schwinger model with a $\theta$-term as a benchmark model, we studied the dependence of the mass shift on the lattice volume $N/\sqrt{x}$, the lattice spacing $ag$, the $\theta$-parameter, and the Wilson parameter $r$. We found that the mass shift shows a strong volume dependence for small values of $N/\sqrt{x}$, but this dependence becomes negligible upon reaching sufficiently large volumes, $N/\sqrt{x} \gtrsim 40$. The mass shift also strongly depends on the lattice spacing and is antisymmetric in the Wilson parameter, as expected. Given that the chiral anomaly is not exact on the lattice, the mass shift has a weak dependence on $\theta$, which vanishes as $ag$ goes to zero. We demonstrated that incorporating the mass shift leads to faster convergence of the continuum extrapolation and results in much smaller errors. In the future, our method could be applied to various lattice calculations in the Hamiltonian formalism, such as the computation of the Schwinger boson mass and classical or quantum computations of other lattice field theories in 1+1 and higher dimensions.

\acknowledgments
L.F.\ is partially supported by the U.S.\ Department of Energy, Office of Science, National Quantum Information Science Research Centers, Co-design Center for Quantum Advantage (C$^2$QA) under contract number DE-SC0012704, by the DOE QuantiSED Consortium under subcontract number 675352, by the National Science Foundation under Cooperative Agreement PHY-2019786 (The NSF AI Institute for Artificial Intelligence and Fundamental Interactions, \url{http://iaifi.org/}), and by the U.S.\ Department of Energy, Office of Science, Office of Nuclear Physics under grant contract numbers DE-SC0011090 and DE-SC0021006.
S.K.\ acknowledges financial support from the Cyprus Research and Innovation Foundation under projects ``Future-proofing Scientific Applications for the Supercomputers of Tomorrow (FAST)'', contract no.\ COMPLEMENTARY/0916/0048, and “Quantum Computing for Lattice Gauge Theories (QC4LGT)”, contract no. EXCELLENCE/0421/0019.

\appendix

\section{Staggered Fermions}
\label{staggered_appendix}

\begin{figure}
    \centering
    \includegraphics[width=0.5\textwidth]{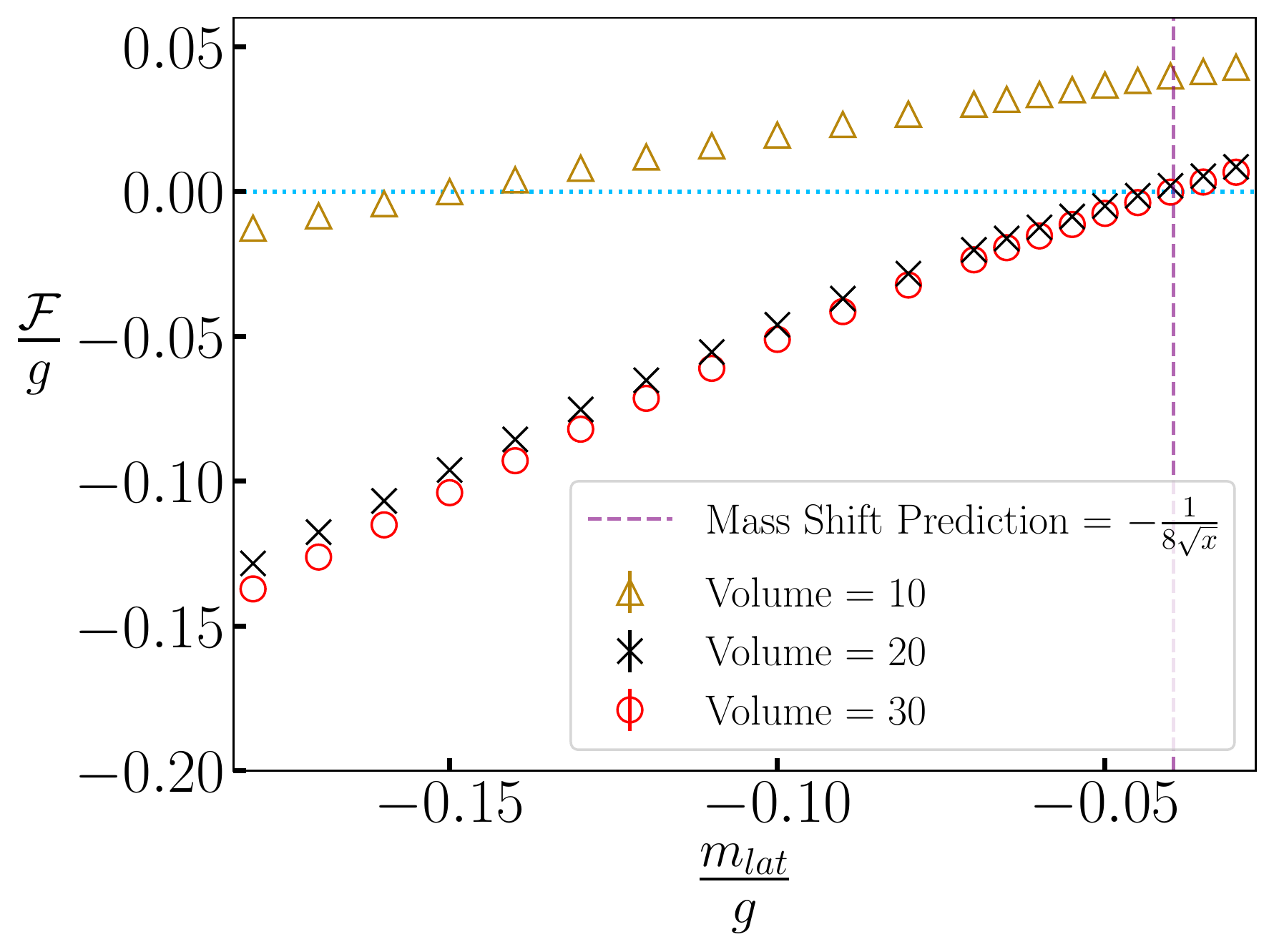}
    \caption{Electric field density $\mathcal{F}/g$ as a function of the lattice mass $m_\text{lat}/g$ using Kogut-Susskind staggered fermions, for different volumes $N/\sqrt{x}$. The vertical purple dashed line is the theoretical prediction of Ref.~\cite{chiral_dempsey_staggered}. While this prediction was derived using periodic boundary conditions, our data is for open boundary conditions and therefore only matches the prediction for sufficiently large volumes.}
    \label{staggered_mass_shift_plot}
\end{figure}

Our new method of computing the mass shift is not only applicable to Wilson fermions but also to staggered fermions~\cite{mass_spectrum_staggered}, as we briefly discuss in this appendix. 
For the one-flavour Schwinger model with a $\theta$-term and staggered fermions, Ref.~\cite{chiral_dempsey_staggered} recently showed that this lattice model receives an additive mass renormalization given by $m_r/g = m_{\text{lat}}/g + 1/(8\sqrt{x})$. This prediction was derived by enforcing a discrete spurious chiral symmetry on the staggered-fermion lattice with periodic boundary conditions and $m_r/g = 0$, which is given by a translation of one site followed by a shift of $\theta$ by $\pi$. Using our new method of numerically determining the mass shift, we provide a finite-volume test of this prediction for OBC and find agreement for sufficiently large lattice volumes, $N/\sqrt{x}\gtrsim 30$, as shown in Fig.\ \ref{staggered_mass_shift_plot}. 

\bibliographystyle{JHEP}
\bibliography{Papers}

\end{document}